\documentclass[prd,english,aps,showpacs,nofootinbib,preprint,eqsecnum]{revtex4-1}
\usepackage[T1]{fontenc}
\usepackage[latin9]{inputenc}
\usepackage{amssymb}
\usepackage{graphicx,color,amsmath,amsxtra}
\usepackage{esint}

\makeatletter

\providecommand{\tabularnewline}{\\}

\@ifundefined{textcolor}{}
{%
 \definecolor{BLACK}{gray}{0}
 \definecolor{WHITE}{gray}{1}
 \definecolor{RED}{rgb}{1,0,0}
 \definecolor{GREEN}{rgb}{0,1,0}
 \definecolor{BLUE}{rgb}{0,0,1}
 \definecolor{CYAN}{cmyk}{1,0,0,0}
 \definecolor{MAGENTA}{cmyk}{0,1,0,0}
 \definecolor{YELLOW}{cmyk}{0,0,1,0}
 }

\usepackage{enumerate}
\usepackage{hhline}

\makeatother

\usepackage{babel}
\begin{document}

\title{Gravitational Waves in Viable $f(R)$ Models}

\author{Louis Yang}

\email{louis.lineage@msa.hinet.net}

\selectlanguage{english}%

\author{Chung-Chi Lee}

\email{g9522545@oz.nthu.edu.tw}

\selectlanguage{english}%

\author{Chao-Qiang Geng}

\email{geng@phys.nthu.edu.tw}

\selectlanguage{english}%

\affiliation{Department of Physics, National Tsing Hua University, Hsinchu 300,
Taiwan\\
Physics Division, National Center for Theoretical Sciences, Hsinchu
300, Taiwan}

\date{\today}
\begin{abstract}
We study gravitational waves in viable $f(R)$ theories under a non-zero
background curvature. In general, an $f(R)$ theory contains an extra
scalar degree of freedom corresponding to a massive scalar mode of
gravitational wave. For viable $f(R)$ models, since there always
exits a de-Sitter point where the background curvature in vacuum is
non-zero, the mass squared of the scalar mode of gravitational wave
is about the de-Sitter point curvature $R_{d}\sim10^{-66}eV^{2}$.
We illustrate our results in two types of viable $f(R)$ models: the
exponential gravity and Starobinsky models. In both cases, the mass
will be in the order of $10^{-33}eV$ when it propagates in vacuum.
However, in the presence of matter density in galaxy, the scalar mode
can be heavy. Explicitly, in the exponential gravity model, the mass
becomes almost infinity, implying the disappearance  of the scalar
mode of gravitational wave, while the Starobinsky model gives the
lowest mass around $10^{-24}eV$, corresponding to the lowest frequency of $10^{-9}$~Hz,
which may be detected by the current and future gravitational wave probes, such as LISA and ASTROD-GW.
\end{abstract}


\maketitle

\section{Introduction}

The accelerating expansion of our universe has been supported  by various
cosmological observations, such as type Ia supernovae~\cite{Riess1998a,Perlmutter1999a},
large scale structure~\cite{Tegmark2004a,Seljak2005a}, cosmic microwave
background radiation~\cite{Spergel2003,Spergel2007} and weak lensing
\cite{Jain2003}. In order to explain this late time acceleration~\cite{DDE},
one can either introduce dark energy, a new form of matter, or modify
Einstein\textquoteright{}s general relativity, i.e., the modification
of gravity. One simple way to modify general relativity is to promote
the Ricci scalar $R$ in the Einstein-Hilbert action into an $f(R)$
function, which is the so-called $f(R)$ theory~\cite{Carroll2004,Nojiri2006c,Faraoni2008d,DeFelice2010b,Sotiriou2010}.
A viable model of $f(R)$ can generate a late-time accelerating expansion
of our universe, have the radiation-dominated stage followed by the
matter-dominated one~\cite{Amendola2007c,Nojiri:2006gh}, and be consistent with
the solar-system constraint under chameleon mechanism~\cite{Mota:2003tc,Khoury2004b,Faulkner2007,Hu2007,Navarro2007,Capozziello2008h,Tsujikawa2008,Tsujikawa2008b}.

In~\cite{Chiba2003a}, Chiba showed that an $f(R)$ model will allow
a new scalar degree of freedom. This
corresponds to a new scalar mode of gravitational wave besides the
ordinary tensor one of  general relativity. 
This new scalar mode will be massive and propagate
as a longitudinal polarization. Various discussions and predictions
about this extra scalar mode of gravitational wave have been given in  the 
literature~\cite{Nojiri:2003ni,Corda2007a,Capozziello2008,Corda2008d,Corda2008e,Corda2010d,Berry2011,Naf2011}.
However, most of them were concentrated on either quadratic or inverse-curvature
type of $f(R)$ models, which is highly restricted by the observational
results~\cite{Amendola2007,Amendola2007a,Navarro2007,Sawicki2007,Starobinsky2007,DeFelice2010b}.
In the article, we will study gravitational wave in viable $f(R)$
models. 

The mass of the scalar mode of gravitational wave in viable $f(R)$
models could be quite different from quadratic and inverse-curvature
ones. In vacuum, it will be of the order of the Hubble constant
because all viable $f(R)$ models need to have de Sitter points which
have a non-zero background curvature $R_{d}$ about square of the
Hubble constant. However, when gravitational wave propagates in the galaxy region,
the local density of dark matter and baryonic matter will contribute
a tremendous background curvature although it is still much smaller
then curvature generated by star. This background curvature might
make some viable $f(R)$ models return to the ordinary GR very fast.
Therefore, the scalar mode will become extremely massive in these
cases and, hence, prevent the observable propagation of the fifth
force in our galaxy and solar system. On the other hand, viable $f(R)$
theories pass the solar-system constraint by means of chameleon mechanism,
which do not require very heavy scalar modes by using the thin-shell
argument~\cite{Faulkner2007,Capozziello2008h}. In these models, the
scalar modes may still be detectable. 

This paper is organized as follows. In Sec. \ref{sec:2}, we
review the field equations and linearizations in $f(R)$ theories
and demonstrate the modifications of gravitational waves. We
apply the analysis on two explicit viable $f(R)$ models: exponential
gravity and Starobinsky ones in Sec. \ref{sec:3}. The results and
discussions on the scalar modes of gravitational waves in the inner
galaxy region is presented in Sec. \ref{sec:4}. Conclusions
are given in Sec. \ref{sec:5}. We use units of $k_{B}=c=\hslash=1$
and the gravitational constant $G=M_{Pl}^{-2}$ with the Planck mass
of $M_{Pl}=1.22089\times10^{19}GeV$.

\section{Gravitational Waves in $f(R)$ Theory\label{sec:2}}

\subsection{$f(R)$ Gravity}

We start by considering a general Einstein-Hilbert action

\begin{equation}
S=\frac{1}{2\kappa^{2}}\int d^{4}x\sqrt{-g}f(R)+S_{m}(g_{\mu\nu},\Psi_{\mu\nu}),\label{eq:action}
\end{equation}
where $f(R)$ is an arbitrary function of the Ricci scalar $R$, $S_{m}$
is the action of the matter part and $\kappa^{2}\equiv8\pi G$. In
the metric formalism, we vary the action (\ref{eq:action}) with respect
to $g_{\mu\nu}$, and the modified Einstein field equation can be
obtained as
\begin{equation}
f'(R)R_{\mu\nu}-\frac{1}{2}f(R)g_{\mu\nu}+\left(g_{\mu\nu}\square-\nabla_{\mu}\nabla_{\nu}\right)f'(R)=\kappa^{2}T_{\mu\nu},\label{eq:field eq}
\end{equation}
where a prime denotes the derivative with respect to $R$, $\nabla_{\mu}$
is the covariant derivative and $\square=g^{\mu\nu}\nabla_{\mu}\nabla_{\nu}$
is the d'Alembert operator. The trace of the field equation (\ref{eq:field eq})
gives

\begin{equation}
f'(R)R-2f(R)+3\square f'(R)=\kappa^{2}T,\label{eq:trace field eq}
\end{equation}
where $T=g^{\mu\nu}T_{\mu\nu}=-\rho+3a^{2}P$ is the trace of the
matter energy-momentum tensor, and $a$ is the scale factor.

For $f(R)$, the de Sitter stage is a vacuum solution with a positive
constant background curvature $R_{d}$, which is assumed to be homogeneous
and static. Consequently, one has 
\begin{equation}
\nabla_{\mu}f'(R_{d})=0\quad\mathrm{and}\quad f'(R_{d})R_{d}=2f(R_{d}).\label{eq:de Sitter R cond}
\end{equation}
Moreover, from Eq. (\ref{eq:field eq}), the Ricci tensor satisfies
$R_{\mu\nu}|_{R_{d}}=g_{\mu\nu}R_{d}/4$.

\subsection{The Weak-field Approximation}

In order to investigate gravitational wave in $f(R)$ theories, we
need to study the linearized theory of $f(R)$ gravity. Consider a
small perturbation from the FRW metric: 
\begin{equation}
g_{\mu\nu}=\overline{g}_{\mu\nu}+h_{\mu\nu},
\end{equation}
where $\left|h_{\mu\nu}\right|\ll1$ is the perturbation and $\overline{g}_{\mu\nu}=diag(-1,a^{2},a^{2},a^{2})$
is the FRW background metric. If the evolution of the system is much
shorter than Hubble time, we can approximate the background spacetime
to be nearly the Minkowski one with $\overline{g}_{\mu\nu}\approx\eta_{\mu\nu}=diag(-1,1,1,1)$.
We keep the theory to be the first order in $h_{\mu\nu}$ and neglect
terms higher than $\mathcal{O}\left(h^{2}\right)$. Thus, the inverse
of the metric tensor is given by
\begin{equation}
g^{\mu\nu}=\overline{g}^{\mu\nu}-h^{\mu\nu}.
\end{equation}
Note that all indices are raising and lowering by the background metric
$\overline{g}_{\mu\nu}$. In the metric formalism, the perturbation
of connection is

\begin{equation}
\delta\Gamma_{\alpha\beta}^{\gamma}=\frac{1}{2}\overline{g}^{\gamma\mu}\left(\partial_{\beta}h_{\alpha\mu}+\partial_{\alpha}h_{\mu\beta}-\partial_{\mu}h_{\alpha\beta}-2h_{\mu\nu}\bar{\Gamma}_{\alpha\beta}^{\nu}\right),
\end{equation}
where $\bar{\Gamma}_{\alpha\beta}^{\nu}$ is the unperturbed connection.
The only non-vanishing components of $\bar{\Gamma}_{\alpha\beta}^{\nu}$
are
\begin{equation}
\bar{\Gamma}_{j0}^{i}=\bar{\Gamma}_{0j}^{i}=H\delta_{j}^{i}\quad\mathrm{and}\quad\bar{\Gamma}_{ij}^{0}=a^{2}H\delta_{ij},
\end{equation}
where $H\equiv\dot{a}/a$ is the Hubble constant. The deviation of
the Ricci tensor from the background curvature is~\cite{Weinberg}
\begin{equation}
\delta R_{\alpha\beta}=\partial_{\mu}\delta\Gamma_{\alpha\beta}^{\mu}-\partial_{\beta}\delta\Gamma_{\alpha\mu}^{\mu}+\bar{\Gamma}_{\alpha\beta}^{\nu}\delta\Gamma_{\nu\mu}^{\mu}+\bar{\Gamma}_{\nu\mu}^{\mu}\delta\Gamma_{\alpha\beta}^{\nu}-\bar{\Gamma}_{\alpha\mu}^{\nu}\delta\Gamma_{\nu\beta}^{\mu}-\bar{\Gamma}_{\nu\beta}^{\mu}\delta\Gamma_{\alpha\mu}^{\nu}.
\end{equation}

\subsection{The Scalar Mode $h_{f}$}

The different between gravitational waves in $f(R)$ and general relativity
is that it contains an extra scalar degree of freedom in $f(R)$.
This comes from the non-vanishing trace of the field equation. Eq.
(\ref{eq:trace field eq}) can be viewed as equation of motion for
a scalar field $\Phi$. By the identifications~\cite{Starobinsky1980,Capozziello2007f}
\begin{equation}
\Phi\rightarrow f'(R)\quad\mathrm{and}\quad\frac{dV_{eff}}{d\Phi}\rightarrow\frac{2f(R)-f'(R)R-\kappa^{2}\rho}{3},
\end{equation}
we obtain the Klein-Gordon equation for the scalar field $\Phi$:
\begin{equation}
\square\Phi=\frac{dV_{eff}}{d\Phi}.
\end{equation}
In order to have a stable perturbation of spacetime, we must require
the background scalar $\Phi_{0}$ to stay at the stable minimum of
the effective potential $V_{eff}$, i.e., 
\begin{equation}
\frac{dV_{eff}}{d\Phi}=0\label{eq:dVeff=00003D0}
\end{equation}
 and 
\begin{equation}
\frac{d^{2}V_{eff}}{d\Phi^{2}}>0.\label{eq:ddVeff>0}
\end{equation}
In vacuum, Eq. (\ref{eq:dVeff=00003D0}) just gives us the condition
for the de-Sitter point curvature (\ref{eq:de Sitter R cond}), while
Eq. (\ref{eq:ddVeff>0}) requires the mass of the scalar mode to be
positive.

Perturbing the trace of the field equation (\ref{eq:trace field eq})
with a nonzero constant background curvature $R_{0}$ which satisfies
Eq. (\ref{eq:dVeff=00003D0}) yields 
\begin{equation}
3\square\delta f'+R_{0}\delta f'+f'(R_{0})\delta R-2\delta f=0.
\end{equation}
Using the relations $\delta f=f'(R_{0})\delta R$ and $\delta f'=f''(R_{0})\delta R$,
we obtain the massive wave equation for the scalar mode~\cite{Capozziello2008,Corda2010d}
\begin{equation}
\square h_{f}=m_{s}^{2}h_{f},\label{eq:scalar wave eq}
\end{equation}
where $h_{f}\equiv\delta f'/f'(R_{0})$ is the field of the scalar
mode and 
\begin{equation}
m_{s}^{2}=\frac{1}{3}\left(\frac{f'(R_{0})}{f''(R_{0})}-R_{0}\right)\label{eq:mass of scalar mode}
\end{equation}
is the mass squared of it. Note that $m_{s}^{2}=V''_{eff}(\Phi)$
\cite{Hu2007}. For any viable $f(R)$ model, the condition $m_{s}^{2}>0$
is needed for the stability of the cosmological perturbation and to
prevent the field from being a tachyon~\cite{Starobinsky1980,Faraoni2006b,Amendola2007,Sawicki2007,Song2007a,Starobinsky2007}.

For the FRW metric, Eq. (\ref{eq:scalar wave eq}) should be expressed
as
\begin{equation}
\left(-\partial_{0}^{2}+\frac{\partial_{i}^{2}}{a^{2}}-3H\partial_{0}\right)h_{f}=m_{s}^{2}h_{f},
\label{2.17}
\end{equation}
where the term $-3H\partial_{0}$ gives a damping factor caused by
the expansion of the universe. 
To illustrate the solution of Eq.~(\ref{2.17}), we take the de Sitter universe with a constant $H$. In this case,
the solution  is a damped plane
wave
\begin{equation}
h_{f}=A(\vec{k})e^{-\frac{3}{2}Ht}exp\left(iq^{\mu}x_{\mu}\right),
\end{equation}
where $q^{\mu}\equiv(\omega_{m},\vec{k})$, $\omega_{m}=\sqrt{\vec{k}^{2}/a^{2}+m_{s}^{2}-\frac{9}{4}H^{2}}$
is the angular frequency and $A(\vec{k})$ is the amplitude. 
For simplicity and without loss of generality, we take $a=1$ and neglect the damping effect as $\vec{k}^{2}/a^{2}\gg H^2$.
As a result, Eq. (\ref{eq:scalar wave eq}) leads to a simple plane wave
solution 
\begin{equation}
h_{f}=A(\vec{p})exp\left(iq^{\mu}x_{\mu}\right),
\end{equation}
with $\omega_{m}=\sqrt{\vec{k}^{2}+m_{s}^{2}}$. We can see that $m_{s}$
is the cutoff frequency of the scalar mode of gravitational wave.
For $\omega_{m}<m_{s}$,
the wave vector becomes imaginary. The waveform is an exponential decay in distance, i.e.,  
$h_f\propto \exp (-\vec{k}\cdot \vec{x})$. Thus, the scalar will not propagate in space below the cutoff frequency.
The massive scalar mode will not propagate at the speed of light with
the group-velocity 
\begin{equation}
v_{g}=\frac{\left|\vec{k}\right|}{\omega_{m}}=\frac{\sqrt{\omega_{m}^{2}-m_{s}^{2}}}{\omega_{m}}.
\end{equation}

\subsection{The Tensor Mode $h_{\mu\nu}^{T}$}

Perturbing the field equation (\ref{eq:field eq}) under the de-Sitter
curvature $R_{d}$ leads to 
\begin{equation}
f'(R_{d})\delta R_{\mu\nu}+R_{\mu\nu}|_{R_{d}}\delta f'-\frac{1}{2}\overline{g}_{\mu\upsilon}\delta f-\frac{1}{2}h_{\mu\nu}f(R_{d})+\left(\overline{g}_{\mu\nu}\square-\nabla_{\mu}\nabla_{\nu}\right)\delta f'=0.\label{eq:perturb field eq}
\end{equation}
Using $\delta f=\frac{f'}{f''}\delta f'$, $R_{\mu\nu}|_{R_{d}}=\overline{g}_{\mu\nu}R_{d}/4$
and the condition for the de-Sitter stage curvature (\ref{eq:de Sitter R cond}),
Eq. (\ref{eq:perturb field eq}) becomes
\begin{equation}
\delta R_{\mu\nu}+\frac{1}{4}\overline{g}_{\mu\nu}R_{d}h_{f}-\frac{1}{2}\overline{g}_{\mu\nu}\frac{f'}{f''}h_{f}-\frac{1}{4}R_{d}h_{\mu\nu}+\left(\overline{g}_{\mu\nu}\square-\nabla_{\mu}\nabla_{\nu}\right)h_{f}=0.\label{eq:perturb field eq 2}
\end{equation}
The total perturbation of metric can be decomposed into the tensor
part $h_{\mu\nu}^{T}$ and scalar part $h_{\mu\nu}^{S}$~\cite{Weinberg}
in the way that
\begin{equation}
h_{\mu\nu}=h_{\mu\nu}^{T}+h_{\mu\nu}^{S},\label{eq:hmn}
\end{equation}
where $h_{\mu\nu}^{S}=b\overline{g}_{\mu\nu}h_{f}$ and $b$ is an
unknown factor which will be determined later. Note that we require
the tensor mode $h_{\mu\nu}^{T}$ to be traceless because we do not
want it to give any contribution to the purturbation of Ricci scalar
$\delta R$ and couple to the scalar mode $h_{f}$. Since the theory
is linearized, the perturbation of Ricci curvature tensor can also
be seperated into two parts
\begin{equation}
\delta R_{\mu\nu}=\delta R_{\mu\nu}^{T}+\delta R_{\mu\nu}^{S},\label{eq:dRmn}
\end{equation}
where $\delta R_{\mu\nu}^{T}$ and $\delta R_{\mu\nu}^{S}$ represent
the perturbations contributed from the tensor mode $h_{\mu\nu}^{T}$
and scalar mode $h_{\mu\nu}^{S}$, respectively. In the de Sitter
universe, we have $\partial_{0}H=0$ and $R_{d}=12H^{2}$. Thus, $\delta R_{\mu\nu}^{S}$
can be written as~\cite{Weinberg}
\begin{equation}
\delta R_{\mu\nu}^{S}=-b\left(\frac{1}{2}\overline{g}_{\mu\nu}\square+\nabla_{\mu}\nabla_{\nu}\right)h_{f}.
\end{equation}
Inserting this into Eq. (\ref{eq:perturb field eq 2}), we obtain
\begin{equation}
\delta R_{\mu\nu}^{T}-\frac{1}{4}R_{d}h_{\mu\nu}^{T}+\left[\left(1-\frac{b}{2}\right)\overline{g}_{\mu\nu}\square-\left(1+b\right)\nabla_{\mu}\nabla_{\nu}+\frac{1}{4}\left(1-b\right)\overline{g}_{\mu\nu}R_{d}-\frac{1}{2}\overline{g}_{\mu\nu}\frac{f'}{f''}\right]h_{f}=0.
\end{equation}
Since the wave equation for the scalar mode (\ref{eq:scalar wave eq}) does not involve any off-diagonal term, and we do not want the coupling between scalar and tensor modes, the term $\nabla_{\mu}\nabla_{\nu}h_{f}$ should not appear here.
To cancel the $\nabla_{\mu}\nabla_{\nu}$ term, we must pick $b=-1$
\cite{Capozziello2008,Corda2010d,Berry2011}. This gives
\begin{equation}
\delta R_{\mu\nu}^{T}-\frac{1}{4}R_{d}h_{\mu\nu}^{T}+\frac{1}{2}\overline{g}_{\mu\nu}\left(3\square+R_{d}-\frac{f'}{f''}\right)h_{f}=0,
\end{equation}
where the terms in parentheses vanish by using the wave equation for
the scalar mode (\ref{eq:scalar wave eq}). 
It is clear that the extra
scalar degree of freedom in $f(R)$ can totally decouple from the
ordinary tensor mode: 
\begin{equation}
\delta R_{\mu\nu}^{T}-\frac{1}{4}R_{d}h_{\mu\nu}^{T}=0.\label{eq:tensor EoM}
\end{equation}

Similar to gravitational  wave in GR, we assume the tensor mode to
be divergenceless, which means that it has to satisfy the Lorenz gauge
condition, i.e., $\partial^{\mu}h_{\mu\nu}^{T}=0$~\cite{Capozziello2008,Berry2011},
and be transverse, $h_{\mu0}^{T}=h_{0\mu}^{T}=0$. The perturbation
of Ricci tensor for the tensor mode can be simplified in this case
as
\begin{equation}
\delta R_{\mu\nu}^{T}=\frac{-1}{2}\left(-\partial_{0}^{2}+\frac{\partial_{i}^{2}}{a^{2}}+H\partial_{0}-4H^{2}\right)h_{\mu\nu}^{T}.\label{eq:RijT perturb}
\end{equation}
We further define $h_{ij}^{T}=a^{2}\mathcal{H}_{ij}$ to absorb the
effect of expansion of the universe~\cite{Dodelsonb}, and choose
the tensor mode to propagate in the z direction with
\begin{equation}
\mathcal{H}_{ij}=\left(\begin{array}{ccc}
h_{+} & h_{\times} & 0\\
h_{\times} & -h_{+} & 0\\
0 & 0 & 0
\end{array}\right),
\end{equation}
where $h_{+}$ and $h_{\times}$ are the plus and cross polarizations,
respectively. The first and second time derivatives of $h_{ij}^{T}$
can be expressed as
\begin{equation}
\partial_{0}h_{ij}^{T}=a^{2}\left(\partial_{0}+2H\right)\mathcal{H}_{ij}\quad\mathrm{and}\quad\partial_{0}^{2}h_{ij}^{T}=a^{2}\left(\partial_{0}^{2}+4H\partial_{0}+4H^{2}\right)\mathcal{H}_{ij}.
\end{equation}
Inserting these and Eq. (\ref{eq:RijT perturb}) into (\ref{eq:tensor EoM}),
we obtain the wave equation for the tensor mode 
\begin{equation}
\left(-\partial_{0}^{2}+\frac{\partial_{i}^{2}}{a^{2}}-3H\partial_{0}\right)\mathcal{H}_{ij}=0\label{eq:tensor wave eq}
\end{equation}
or
\begin{equation}
\square h_{\alpha}=0,\quad\alpha=+,\times.
\end{equation}
Note that Eq. (\ref{eq:tensor wave eq}) is equivalent to Eq. (5.1.53)
in Ref.~\cite{Weinberg} and Eq. (5.61) in Ref.~\cite{Dodelsonb}.
Clearly,  the tensor mode is exactly the same as
that in GR when a traceless gravitational wave propagates in a non-zero
de-Sitter curvature $R_{d}$ background. 
In what follows, we will concentrate on the scalar mode of gravitational wave.

\section{Gravitational Waves in Viable $f(R)$ Models\label{sec:3}}

The conditions for a cosmological viable $f(R)$ model include (i)
the positivity of the effective gravitational coupling, (ii) the stability
of cosmological perturbations~\cite{Dolgov2003c,Faraoni2006b,Song2007a,Starobinsky2007},
(iii) the stability of the late-time de-Sitter point~\cite{Tsujikawa2009,Amendola2007c,Muller1988,Faraoni2005},
(iv) the asymptotic behavior to $\Lambda$CDM at the high curvature
regime, (v) the solar system constraint, and (vi) the constraint from
the violation of the equivalence principle~\cite{Hu2007,Faulkner2007,Capozziello2008h}.
The typical examples of the viable $f(R)$ models are Hu-Sawicki~\cite{Hu2007},
Starobinsky~\cite{Starobinsky2007}, Tsujikawa~\cite{Tsujikawa2008}
and exponential gravity models~\cite{Linder2009a}. To illustrate
our results, we will concentrate on the exponential and Starobinsky
models. Our study can be easily extended to other viable models.

Usually a viable $f(R)$ model can use chameleon mechanism to pass
the constraints from the solar-system and equivalence principle~\cite{Hu2007,Faulkner2007,Capozziello2008h}.
In this case, a light-mass scalar is allowed by introducing the thin
shell condition. As a result, the mass of the scalar mode does not
have to be very heavy~\cite{Faulkner2007}.

\subsection{$\Lambda$CDM}

We can take the cosmological constant model or the $\Lambda$CDM model
as a special case of $f(R)$ with 
\begin{equation}
f(R)=R-2\Lambda,
\end{equation}
where $\Lambda$ is the cosmological constant.

In this case, the mass of the scalar mode is infinite, i.e., $m_{s}^{2}=\infty$,
which requires infinite large energy to excite the scalar
mode. Clearly, there is no scalar mode in the $\Lambda$CDM model.
Although the de Sitter curvature $R_{d}$ is not zero, i.e., 
\begin{equation}
R_{d}=4\Lambda\quad\left(\Lambda CDM\right),
\end{equation}
the contribution from $R_{d}$ is negligible because $\Lambda\approx H_{0}^{2}\thickapprox\left(10^{-33}eV\right)^{2},$
where $H_{0}$ is the present Hubble parameter.

\subsection{Exponential Gravity}

The exponential gravity has been studied intensively in the literature
\cite{Zhang2006c,Zhang2007,Tsujikawa2008,Cognola2008a,Linder2009a,Ali2010,Bamba2010f,Yang2010,Elizalde2011}.
The form of $f(R)$ in the exponential gravity model is given by
\begin{equation}
f(R)=R-\beta R_{S}\left(1-e^{-R/R_{S}}\right),\label{eq:exponential gravity}
\end{equation}
 where $R_{S}$ is the characteristic curvature scale and $\beta$
is a model parameter. The viable conditions are satisfied when $\beta>1$
and $R_{S}>0$~\cite{Linder2009a,Bamba2010f}. The feature is that
it is free from the fine tuning problem and it has only one parameter
more than the $\Lambda$CDM model.

Now we will first investigate the de-Sitter curvature $R_{d}$ in
the model. The first and second derivatives of $f(R)$ with respect
to $R$ are 
\begin{equation}
f'(R)=1-\beta e^{-R/R_{S}}\quad\mathrm{and}\quad f''(R)=\frac{\beta}{R_{S}}e^{-R/R_{S}}.
\end{equation}
 According to the condition for the de-Sitter curvature (\ref{eq:de Sitter R cond}),
$R_{d}$ satisfies
\begin{equation}
\left(1-\beta e^{-R_{d}/R_{S}}\right)R_{d}=2R_{d}-2\beta R_{S}\left(1-e^{-R_{d}/R_{S}}\right).\label{eq:Rd for ExpG}
\end{equation}
 Defining $x\equiv R_{d}/R_{S}$, Eq. (\ref{eq:Rd for ExpG}) becomes
\begin{equation}
x=2\beta-\beta e^{-x}\left(x+2\right).\label{eq:beta and x}
\end{equation}
\begin{figure}
\includegraphics{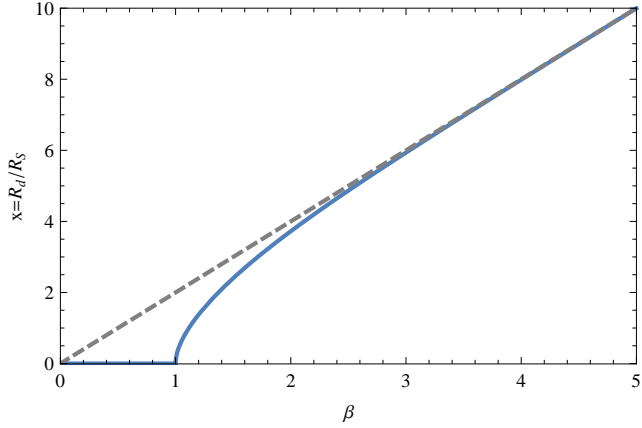}
\caption{Numerical solution of $x=R_{d}/R_{S}$ versus the model parameter
$\beta$ in the exponential gravity model,
where the solid line indicates the exact solution of Eq. (\ref{eq:beta and x})
and the dashed line presents the approximated solution $x=2\beta$. 
\label{fig:x vs beta}}
\end{figure}
The numerical solutions for this equation are shown in Fig. \ref{fig:x vs beta}.
The factor $e^{-x}\left(x+2\right)$ decreases very fast when $\beta>1$,
which is generally required by the viable condition for the exponential
gravity. Therefore, we can obtain the asymptotic solution of $x$
for a large $\beta$:
\begin{equation}
x=2\beta\quad\mathrm{for}\quad\beta\gg1.
\end{equation}

  From Eq. (\ref{eq:mass of scalar mode}), we  derive the mass squared
of the scalar mode in the exponential gravity as
\begin{equation}
m_{s}^{2}=\frac{1}{3}R_{S}\left(\frac{e^{R_{d}/R_{S}}-\beta}{\beta}-\frac{R_{d}}{R_{S}}\right)=\frac{1}{3}R_{S}\left(\frac{1}{\beta}e^{x}-1-x\right).\label{eq:ms in EG}
\end{equation}
\begin{figure*}
\includegraphics{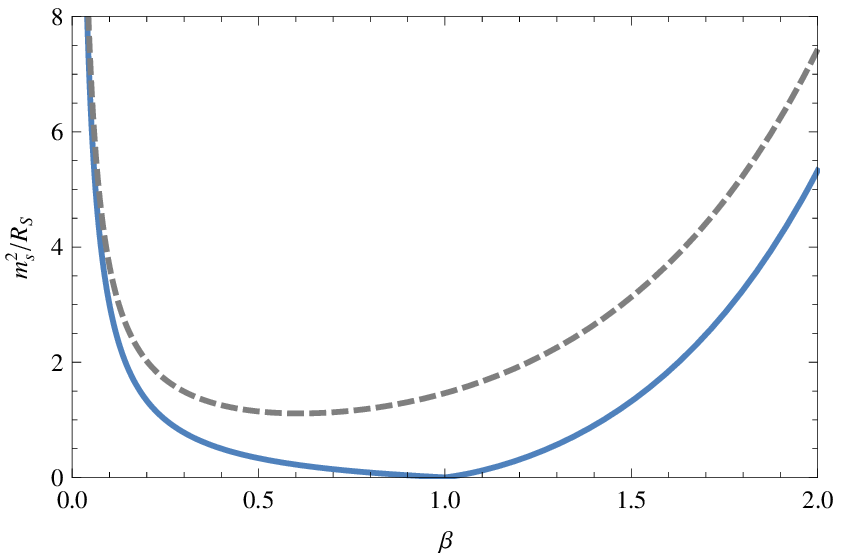}\includegraphics{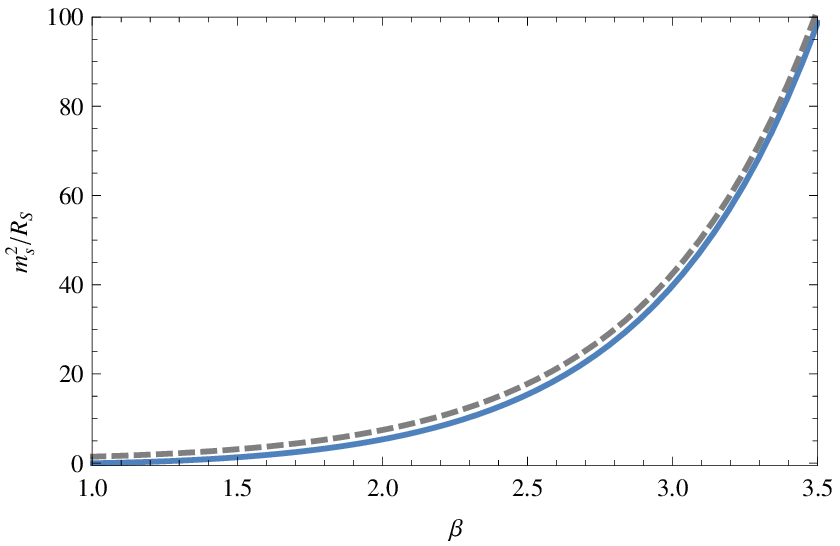}
\caption{$m_{s}^{2}/R_{S}$ versus $\beta$ in the region of $\beta=0$ to
2 (left panel) and $\beta=1$ to 3.5 (right panel) in the exponential
gravity model, where the solid lines
indicate the numerical solution of Eq. (\ref{eq:ms in EG}), $m_{s}^{2}/R_{S}$
is essentially zero when $\beta=1$, and the dashed lines show the
approximated solution $x=2\beta$.\label{fig:ms d Rs vs beta}}
\end{figure*}
The numerical solutions for $m_{s}^{2}/R_{S}$ are presented in Fig.
\ref{fig:ms d Rs vs beta}. Since in the large curvature regime $R/R_{S}\gg1$,
the theory will recover the cosmological constant model, $R_{S}$
is roughly inverse proportional to $\beta$ in the way that
\begin{equation}
\beta R_{S}\cong2\Lambda=9.94\times10^{-66}eV^{2}\label{eq:bRs and L relations}
\end{equation}
with the value of $\Lambda$ obtained from WMAP 7~\cite{Komatsu2011},
SDSS 7~\cite{Percival2010} and SCP Union2 observations~\cite{Amanullah2010}.
Eq. (\ref{eq:ms in EG}) then can be approximated as 
\begin{equation}
m_{s}^{2}\cong\frac{2\Lambda}{3\beta}\left(\frac{1}{\beta}e^{x}-x-1\right).\label{eq:ms of L}
\end{equation}
\begin{figure*}
\includegraphics{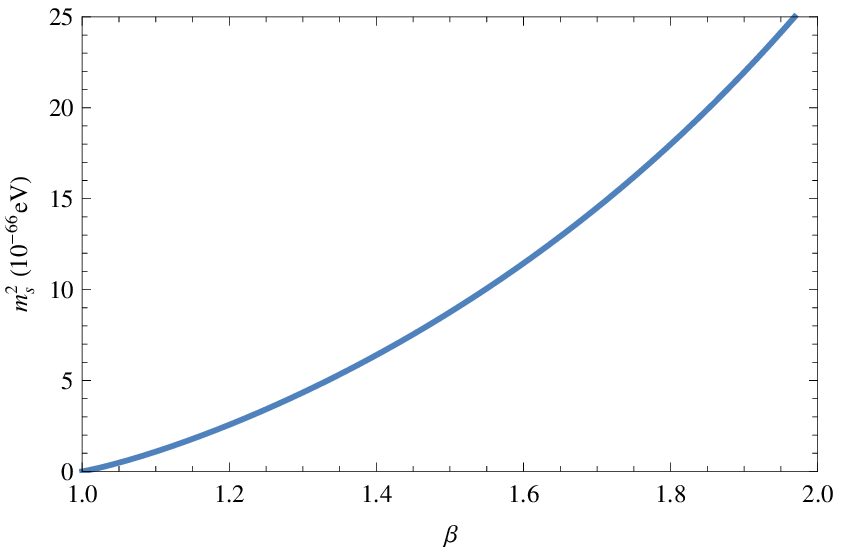}\includegraphics{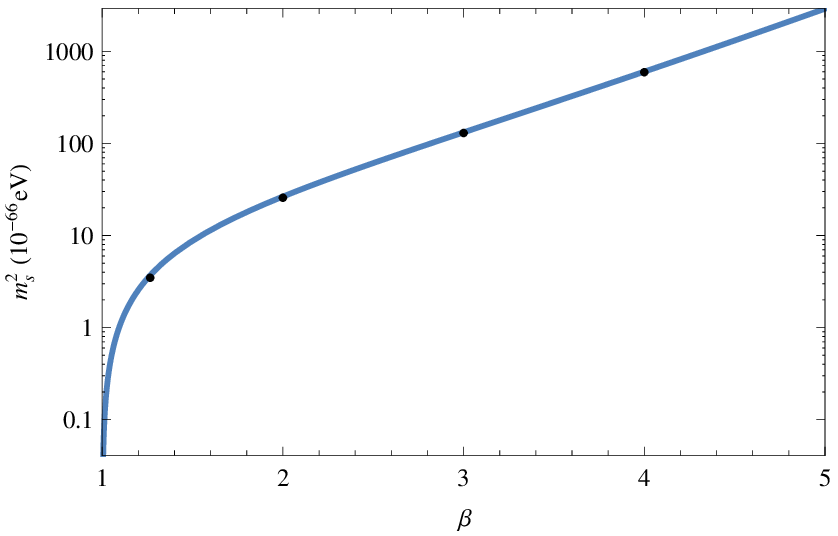}
\caption{$m_{s}^{2}$ versus $\beta$ in the region of $\beta=1$ to 2 (left
panel) and $\beta=1$ to 5 in log scale (right panel) in the exponential
gravity model, where the solid lines
present the approximated numerical solution of $m_{s}^{2}$ obtained
by Eq. (\ref{eq:ms of L}) and the dots in the right panel show the
exact value of $m_{s}^{2}$ presented in Table \ref{tab:ms}. \label{fig:ms}}
\end{figure*}
In Fig. \ref{fig:ms}, we depict the result of the mass squared $m_{s}^{2}$
versus $\beta$ by using Eq. (\ref{eq:ms of L}). We also calculate
the exact $m_{s}$ without any approximation, and the results of $\beta=1.27$,
2, 3 and 4 are shown in Table \ref{tab:ms},
\begin{table}[t]
\caption{Numerical results of the scalar mode mass $m_{s}$ in vacuum with
respect to different $\beta$ in the exponential gravity model. \label{tab:ms}}
\begin{ruledtabular}
\begin{tabular}{cccccc}
$\beta$ & $h$ & $y_{H}^{ini}$ & $\Omega_{m}^{0}$ & $R_{S}\:(10^{-66}eV^{2})$ & $m_{s}\:(10^{-33}eV)$\tabularnewline
\hline 
4 & 0.7050 & 2.618 & 0.2761 & 2.452 & 24.36\tabularnewline
3 & 0.7059 & 2.609 & 0.2758 & 3.263 & 11.39\tabularnewline
2 & 0.7103 & 2.558 & 0.2738 & 4.824 & 5.069\tabularnewline
1.27 & 0.7194 & 2.45 & 0.2701 & 7.39 & 1.86\tabularnewline
\end{tabular}
\end{ruledtabular}
\end{table}
where we have used the values of $R_{S}$ obtained from our previous
result in Ref.~\cite{Yang2010} under the constraints of WMAP 7, SDSS
7 and SCP Union 2 measurements.

For $\beta\gg1$, the mass squared $m_{s}^{2}$ becomes
\begin{equation}
m_{s}^{2}\cong\frac{2\Lambda}{3}\left(\frac{1}{\beta^{2}}e^{2\beta}-2-\frac{1}{\beta}\right).
\end{equation}
When $\beta$ is $\mathcal{O}\left(1\right)$, $m_{s}$ is around $10^{-33}eV$. 
However, the cosmological observations do not
give any significant upper bound on $\beta$. Thus, $m_{s}$ could
be arbitrary large in this case. As $\beta\rightarrow\infty$,corresponding
to the $\Lambda$CDM model with $m_{s}\rightarrow\infty$, the scalar
mode of gravitational wave vanishes.

\subsection{Starobinsky Model}

\begin{figure}
\includegraphics{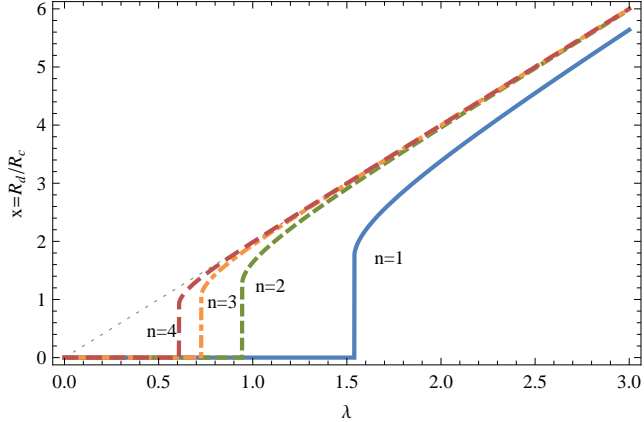}
\caption{Numerical results of the de Sitter curvature $R_{d}$ divided by $R_{c}$
versus $\lambda$ with $n=1$ (solid), 2 (dashed), 3 (dot-dashed)
and 4 (long-dashing), respectively, in the Starobinsky model, where
the dotted line presents the approximation solutions $x=2\lambda$
when $\lambda\gg1$.\label{fig:Star-Rd}}
\end{figure}

In Ref.~\cite{Starobinsky2007}, Starobinsky proposed the following
$f(R)$ form: 
\begin{equation}
f(R)=R-\lambda R_{c}\left(1-\left(1+\frac{R^{2}}{R_{c}^{2}}\right)^{-n}\right),\label{eq:Starobinsky model}
\end{equation}
where $R_{c}$ is roughly the present cosmological density and $\lambda$
and $n$ are positive model parameters. From the solar system constraint
and the bound on the violation of the equivalence principle, one gets
$n>0.9$~\cite{Capozziello2008h}. In Fig. \ref{fig:Star-Rd}, we
present the vacuum curvature $R_{d}$ at the de Sitter stages obtained
by Eq. (\ref{eq:de Sitter R cond}). 
We can see that  $R_{d}/R_{c}\simeq2\lambda$ when $\lambda\gg1$.
Since for $R\gg R_{c}$, the model will restore the $\Lambda$CDM
model, we have $\lambda R_{c}\simeq2\Lambda$ and $R_{d}\simeq4\Lambda=1.99\times10^{-65}eV^{2}$
when $\lambda\gg1$.

\begin{figure}
\includegraphics{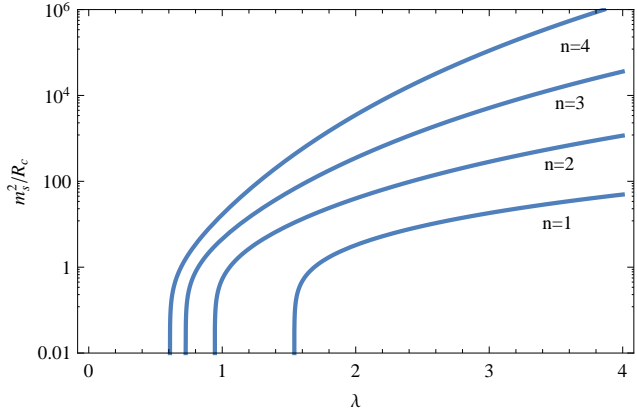}
\caption{$m_{s}^{2}/R_{c}$ versus $\lambda$ in vacuum in the Starobinsky
model. \label{fig:Star-ms of Rc} }
\end{figure}

In Fig.~\ref{fig:Star-ms of Rc}, we depict the numerical solutions
of the mass squared of the scalar mode $m_{s}^{2}$ derived from Eq.
(\ref{eq:mass of scalar mode}).

\section{Gravitational Wave in Inner Galaxy\label{sec:4}}

A scalar mode of gravitational wave with effective zero or non-zero
mass in $f(R)$ is a very different prediction from the ordinary GR.
However, we will show that in the presence of matter density, the
scalar mode might not be able to exist in these viable $f(R)$ models.
Consider a scalar mode of gravitational wave propagating within our
Galaxy halo. The local homogeneous density of dark matter and baryonic
matter is roughly $\rho\approx10^{-24}g/cm^{3}$. If we take this
matter density into our analysis, it will give a large contribution
to the background curvature compared to the vacuum de Sitter curvature.
(The ratio of the matter density $\rho$ to de Sitter curvature $R_{d}$
is about $\kappa^{2}\rho/R_{d}\simeq\kappa^{2}\rho/4\Lambda\approx10^{5}$.)
In this case, the condition for the background curvature $R_{0}$
(\ref{eq:de Sitter R cond}) should be modified as 
\begin{equation}
f'(R_{0})R_{0}=2f(R_{0})-\kappa^{2}\rho,\label{eq:R0 with matter}
\end{equation}
where $R_{0}$ is the background curvature with matter. Note that
for viable $f(R)$ models, the solutions to Eq. (\ref{eq:R0 with matter})
can be approximated as $R_{0}\simeq\kappa^{2}\rho$ at the high curvature
regime. 

In the case of the exponential gravity (\ref{eq:exponential gravity}),
Eq. (\ref{eq:R0 with matter}) gives 
\begin{equation}
x=2\beta+r-\beta e^{-x}\left(x+2\right),\label{eq:beta and x and r}
\end{equation}
where $x\equiv R_{0}/R_{S}$ and $r\equiv\kappa^{2}\rho/R_{S}$ are
the ratios of the background curvature and matter density to $R_{S}$,
respectively. Since $\beta R_{S}\cong2\Lambda$ from (\ref{eq:bRs and L relations}),
we find that the solution of Eq. (\ref{eq:beta and x and r}) is extremely
large,
\begin{equation}
x\simeq r\simeq\frac{\kappa^{2}\rho}{R_{d}/2\beta}\simeq2\times10^{5}\beta,
\end{equation}
which just leads to $R_{0}\simeq\kappa^{2}\rho$. Thus, in the exponential
gravity, the mass of the scalar mode will become an extreme in the
galaxy region: 
\begin{equation}
m_{s}\approx\sqrt{\frac{2\Lambda}{3\beta^{2}}e^{2\times10^{5}\beta}}\approx\infty.
\end{equation}
The corresponding cutoff frequency $\omega_{m}$ is also infinite.
As a result, it is almost impossible to detect this scalar mode within
our Galaxy under the exponential gravity scenario. Moreover, for any
source that is massive enough to generate gravitational waves, we
expect them to lay in the region with density higher than the
baryonic/dark matter density $10^{-24}g/cm^{3}$. Therefore, the scalar
mode will not have the chance to propagate from the source in the
exponential gravity. 

In the case of the Starobinsky model (\ref{eq:Starobinsky model}),
the situation is quite different. The scalar mode of
gravitational wave can have a light mass in the galaxy region. The
minimum bound of the scalar mode mass is $m_{s}\gtrsim10^{-24}eV$
when $\rho=10^{-24}g/cm^{3}$. The corresponding cutoff frequency
is quite small $f_{m}\gtrsim10^{-9}$ Hz. This feature will allow
the propagation of the scalar mode inside the galaxy. Hence, detecting
the scalar mode in the Starobinsky model will be possible. 
\begin{figure}
\includegraphics{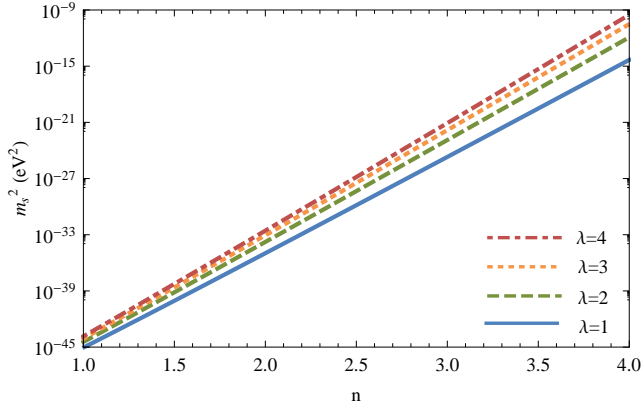}
\caption{$m_{s}^{2}$ versus $n$ in the Starobinsky model with matter density
$\rho=10^{-24}g/cm^{3}$ and $\lambda R_{c}\cong2\Lambda$. \label{fig:Star-ms with matter}}
\end{figure}
In Fig.~\ref{fig:Star-ms with matter}, we depict the mass squared
versus the model parameter $n$ with different fixed values of $\lambda$,
where we have used $\lambda R_{c}\cong2\Lambda$. The scalar mode
can still be very heavy when the index $n$ goes to a large value,
but the mass dependence on the parameter $\lambda$ is not quite significant.

\section{Conclusions\label{sec:5}}

We have discussed gravitational waves in viable $f(R)$ theories.
Using the weak field approximation on the field equation, we have
confirmed that $f(R)$ will give an extra massive scalar mode besides
the ordinary tensor mode in the standard GR. We have explicitly investigated
the situations of the extra scalar mode of gravitational wave in the
exponential gravity and Starobinsky models of the viable $f(R)$ gravity
theories. In vacuum, we have shown that the typical mass squared of
the scalar mode is in the order of the de-Sitter curvature $m_{s}^{2}\sim R_{d}\approx10^{-66}eV^{2}$
in both models. 

However, in the galaxy region, the situations will be different if
we consider the background curvature by the presence of the galactic
density. The small matter density $\rho=10^{-24}g/cm^{3}$ is still
$10^{5}$ larger than the de-Sitter curvature $R_{d}$ in both models.
In the exponential gravity, since the mass of the scalar mode in galaxy
is undetectable large, it will be unable to measure a gravitational
wave. On the other hand, in the Starobinsky model, the mass can be
much smaller with its lower bound in galaxy being about $10^{-24}eV$
(or $10^{-9}$ Hz). Therefore, it is possible to observe the scalar
mode of gravitational wave in the Starobinsky scenario if there is
an astrophysical source which generates this scalar mode. The extreme
difference of the scalar mode masses also makes the distinction
of $f(R)$ theories from $\Lambda$CDM possible by probing this scalar
mode of gravitational wave.

Recently, there is an underway space-based gravitational wave probing
experiment, the Laser Interferometer Space Antenna 
(LISA)~\cite{LISA}\footnote{Laser Interferometer Space Antenna, http://sci.esa.int/lisa}, 
which is a proposed joint mission of the  European Space Agency (ESA)
and NASA.
It will measure the low-frequency band ($10^{-5}$  to 1~Hz) of
gravitational waves with high signal-to-noise ratio. The gravitational
sources within this band~\cite{Ni} include supermassive black holes, intermediate-mass
black holes, extreme-mass-ratio black hole inspirals, galactic compact
binaries and some primordial gravitational wave sources. It is possible
that these sources also generate the scalar mode of gravitational
wave. Since LISA will be located far from the Earth and other gravitational
sources, the background curvature of it is very low compared
to the ground-based experiments, which allows the propagation of the
scalar mode in some viable $f(R)$. As a result, LISA has a great
chance to direct detect not only the ordinary gravitational wave but
also the clues of the deviation from Einstein's GR by analyzing the
scalar mode behavior of gravitational wave. 
Moreover, if the scalar mode becomes observable, 
many interesting features in the viable $f(R)$ gravity models~\cite{DeFelice2010b,Bamba:2010iy} will appear. 
We note that other gravitational wave probes, such as ASTROD-GW~\cite{ASTROD-GW}
with the sensitivity in the $10^{-7}-10^{-1}$~Hz band, 
 may also detect the extra scalar mode. 
 It is clear that an observation of the scalar mode with a frequency larger than $10^{-8}$~Hz by
 LISA or ASTROD-GW would be associated with the Starobinsky model and is rules out the exponential 
 one.
 Finally, we remark that the scalar mode in a viable $f(R)$ cannot 
 be observed by the ground gravitational searches due to the large background curvature.

\begin{acknowledgments}
We are grateful to Professor Wei-Tou Ni for many useful discussions and reading the
manuscript.
The work was supported in part by National Center of Theoretical Science
and  National Science Council 
(NSC-98-2112-M-007-008-MY3)
of R.O.C.
\end{acknowledgments}

\end{document}